\documentclass{aastex}
\usepackage{spr-astr-addons}
\usepackage{url}
\usepackage{grap
hicx,times}

\RequirePackage{color}

\begin{document}

\title{Viscous-Resistive ADAF with a general Large-Scale Magnetic Field}
\shorttitle{Short article title}

\author{Shahram Abbassi\altaffilmark{1,2}}
\email{abbassi@ipm.ir}
\affil{abbassi@ipm.ir}
\and
\author{Amin Mosallanezhad\altaffilmark{3,4}}
\affil{amin.mosallanezhad@gmail.com}
\altaffiltext{1}{School of Physics, Damghan University, P.O.Box 36715-364, Damghan, Iran}
\altaffiltext{2}{School of Astronomy, Institute for Research in Fundamental Sciences, P.O.Box 19395-5531, Tehran, Iran}
\altaffiltext{3}{Center for Excellence in Astronomy \& Astrophysics (CEAA - RIAAM) - Maragha, IRAN, P. O. Box: 55134 - 441}
\altaffiltext{4ý}{Department of Sciencesý, ýJahrom Universityý, ýP.O.Box 74135-111ý, ýJahromý, ýIran}ýý
\email{abbassi@ipm.ir}

\begin{abstract}
    We have studied the structure of hot accretion flow bathed in a general
large-scale magnetic field. We have considered magnetic parameters
$ \beta_{r,\varphi,z}[=c^2_{r,\varphi,z}/(2c^2_{s})] $, where
$ c^2_{r, \varphi, z} $ are the Alfv\'{e}n sound speeds in three
direction of cylindrical coordinate $ (r,\varphi,z) $. The dominant
mechanism of energy dissipation is assumed to be the magnetic diffusivity
due to turbulence and viscosity in the accretion flow. Also, we adopt
a more realistic model for kinematic viscosity $ (\nu=\alpha c_{s} H) $,
with both $ c_{s} $ and $ H $ as a function of magnetic field. As a result in our
model, the kinematic viscosity and magnetic diffusivity $ (\eta=\eta_{0}c_{s} H ) $
are not constant. In order to solve the integrated equations that govern the behavior
of the accretion flow, a self-similar method is used. It is found that the existence
of magnetic resistivity will
increase the radial infall velocity as well as sound speed and vertical thickness
of the disk. However the rotational velocity of the disk decreases by the increase of
magnetic resistivity. Moreover, we study the effect of three components of global
magnetic field on the structure of the disk. We found out that the radial velocity
and sound speed are Sub-Keplerian for all values of magnetic field parameters, but
the rotational velocity can be Super-Keplerian by the increase of toroidal magnetic
field. Also, Our numerical results show that all components of magnetic field can be
important and have a considerable effect on velocities and vertical thickness of the
disk.
\end{abstract}



\section{Introduction}

The study of advection accretion flows around low luminosity black hole candidates and neutron stars is currently a very active field
of research, both theoretically and observationally (for a review, see Narayan, Mahadevan \& Quataert 1998). Observational evidence for the
existence of low-luminosity black holes at the centre of galaxies and in AGNs (Ho 1999) makes it necessary to revise
theoretical models of the accretion discs. Thereby, the development of the subject of advection-dominated accretion flows (ADAFs) in recent
years has led to global solutions of advection accretion discs around accreting black hole systems and neutron stars. In this case, viscously
generated internal energy is not radiated away efficiently as the gas falls into the potential well of the central mass (as in standard thin disc
models; Shakura \& Sunyaev 1973). Instead, it is retained within the accreting gas and advected radially inward (Narayan \& Yi 1994, hereafter
NY1994) and might eventually be lost into the central object or, in contrast, a considerable portion of it might give rise to wind on to black
holes and neutron stars (Blandford \& Begelman 1999). By definition, ADAFs have very low radiative efficiency and as a consequence they can
be considerably hotter than the gas flow in standard thin disc models (Narayan \& Yi 1995a,b); therefore, they are ultra-dim for their accretion
rates (Phinney 1981; Rees et al. 1982).

A notable problem arises when the accretion disc is threaded by
a magnetic field. In the ADAF models, the temperature of the accretion
disc is so high that the accreting materials are ionized. The magnetic field therefore plays an important role in the dynamics of
accretion flows. Some authors have tried to solve the magnetohydrodynamics
(MHD) equations of magnetized ADAFs analytically. The effect of toroidal magnetic field on the disc were studied for example by Akizuki \& Fukue 2006,
Abbassi et al 2008, 2010 and Faghei 2012. Also Zhand \& Dai 2008 have already considered global magnetic field on the disc.
Ghanbari et al 2007, 2009 present the effect of viscous-resistive in ADAFs bathed in a dipole magnetic field.
Resistive diffusion of magnetic field would be important in some
systems, such as the protostellar discs (Stone et al. 2000; Fleming \& Stone 2003), discs in dwarf
nova systems (Gammie \& Menou 1998), the discs around black holes (Kudoh \&
Kaburaki 1996), and Galactic center (Melia \& Kowalenkov 2001; Kaburaki et al. 2010).
In ADAF models, energy dissipation in the accretion flow can be assumed to be a result of turbulent viscosity and electrical resistivity.
The Magnetic energy density of the flow must be dissipated by ohmic heating with a rate comparable to that of the viscous dissipation (Bisnovatyi-Kogan \&
Ruzmaikan 1974, Bisnovatyi-Kogan, Lovelace 1997). We argue that the ohmic and viscous dissipation must occure as a result of plasma instabilities.
Bisnovatyi-Kogan, Lovelace 1997 point out that ohmic dissipation of magnetic energy density has an important role in heating process if accretion flows with condition of equipartition, $\varepsilon_{mag}\sim\varepsilon_{kin}$, where $\varepsilon_{mag} \& \varepsilon_{kin}$ are the magnetic and kinetic energy density, respectively. In this regard, note that although Narayan \& Yi  (1995) assume an equipartition magnetic field, they do not consider the ohmic dissipation.

Under some conditions, it is important that we consider the effect of resistivity on accretion flows. Kuwabara et al. (2000) showed the results of
global magnetohydrodynamic (MHD) simulations of an accretion flow initially threaded by large-scale poloidal magnetic fields, including the
effects of magnetic turbulent diffusivity. They found the importance of the strength of magnetic diffusivity when they studied it in magnetically
driven mass accretion. They pointed out that the mass outflow depends on the strength of magnetic diffusivity, so that for a highly diffusive
disc, no outflow takes place.

In this paper, we extend the work of Akizuki \& Fukue 2006 and ghanbari et al.
2007, Ghanbari et al 2009 and Faghei 2012 by considering large-scale magnetic field with three components in
cylindrical coordinates $ (r, \varphi, z) $ in a viscose-resistive accretion disc, and
investigate the role of non-constant magnetic diffusivity in the system. Ghanbari, Salehi \& Abbassi (2007) have presented a set
of self-similar solutions for two-dimensional (2D) viscous-resistive
ADAFs in the presence of a dipolar magnetic field of the central
accretor. They have shown that the presence of a magnetic field and
its associated resistivity can considerably change the picture with
regard to accretion flows. The paper is
organized as follows. In section 2 we present the basic magnetohydrodynamics equations, which include
the three components of magnetic field and magnetic resistivity. Self-similar equations are
investigated in section 3 and the summery will come up in section 4.

\section[Basic Equations]{Basic Equations}
    Since we are interested in analyzing the structure of a magnetized ADAF
bathed in a global magnetic field, we consider a magnetic field in the
disc with three components, $ B_{r} $, $ B_{\varphi} $ and $ B_{z} $. we suppose that the gaseous disc
rotating around a compact object of mass $ M_{*} $. Thus, for a steady
axi-symmetric accretion flow, i.e., $ \partial/\partial t = \partial/\partial \varphi =0 $ ,
we can write the standard equations in the cylindrical coordinates
$ (r, \varphi, z) $. We vertically integrated the flow equations and,
all the physical variables become only function of the radial distance $ r $.
Moreover, we neglect the relativistic effects and Newtonian gravity in radial
direction is considered.  For conservation of energy, it is assumed the energy
which is generated due to viscosity and resistivity dissipation are balanced by the radiation
and advection cooling. Under these assumption, we can rewrite the MHD equation as (Zhang \& Dai 2008). So the equation of continuity gives
\begin{equation}\label{continuity}
    \frac{1}{r}\frac{d}{dr}(r \Sigma v_r) = 2 \dot{\rho}H,
\end{equation}
where $ \Sigma $ is the surface density at the cylindrical radius $ r $,
which is define as $ \Sigma = 2 \rho H  $, $ v_r $ the radial infall velocity,
$ \dot{\rho} $ the mass loss rate per unit volume, $ H $ would be the disc half-thickness
and $ \rho $ is the density of the disc.

The equation of motion in the radial direction is
\begin{multline}\label{motion_r}
    v_r \frac{dv_r}{dr} = \frac{v^2_{\varphi}}{r} - \frac{GM_*}{r^2} - \frac{1}{\Sigma}\frac{d}{dr}(\Sigma c^2_{s})- \frac{1}{2\Sigma}\frac{d}{dr}(\Sigma c^2_{\varphi} + \Sigma c^2_{z}) \\
    - \frac{c^2_{\varphi}}{r}
\end{multline}
where $ v_{\varphi} $ is the rotational velocity, $ c_s $ the
isothermal sound speed, which is define as $ c^2_{s} \equiv p_{gas}/\rho $, $ p_{gas} $
being the gas pressure. Here, $ c_r $, $ c_{\varphi} $ and
$ c_z $ are Alfv\'{e}n sound speeds in three direction of cylindrical
coordinate and are defined as
\begin{equation}
c^2_{r, \varphi, z} =\frac{B^2_{r, \varphi, z}}{4 \pi \rho}
\end{equation}

The equation of angular momentum transfer can be written as
\begin{multline}\label{motin_phi}
\frac{v_r}{r}\frac{d}{dr}(r v_{\varphi}) = \frac{1}{r^2 \Sigma}\frac{d}{dr}(r^3 \nu \Sigma \frac{d\Omega}{dr}) + \frac{c_r}{\sqrt{\Sigma}}\frac{d}{dr}(\sqrt{\Sigma} c_{\varphi})\\
+ \frac{c_r c_{\varphi}}{r}
\end{multline}
where $ \Omega(=v_{\varphi}/r) $ is the angular velocity. Also, we assumed
that only the $\varphi$-component of viscose stress tensor is important
which is $ t_{r\varphi}=\mu r d\Omega/dr $ , where $ \mu(= \nu \rho) $
is the viscosity and $ \nu $ is the kinematic coefficient of viscosity.

The hydrostatic balance in vertical direction is integrated to
\begin{equation}\label{motion_z}
    \Omega^2_{K} H^2 - \frac{1}{\sqrt{\Sigma}}c_{r} \frac{d}{dr}(\sqrt{\Sigma} c_{z})H = c^2_{s} + \frac{1}{2}(c^2_{r} + c^2_{\varphi}),
\end{equation}

Now we can write the energy equation considering cooling and heating
processes in an ADAF. Therefore the energy equation will be
\begin{equation}\label{energy}
    \frac{\rho v_{r}}{\gamma -1} \frac{dc^{2}_{s}}{dr} - v_{r} c^{2}_{s} \frac{d\rho}{dr} = Q_{+} - Q_{-}
\end{equation}
where $ \gamma $ is the radio of specific heats. In the right hand side
of energy equation, $ Q_{+} = Q_{vis} + Q_{B} $ is the dissipation rate
by viscosity $ Q_{vis} $ and resistivity $ Q_{B} $, and also $ Q_{-} = Q_{rad} $
represent the energy lose through radiative cooling. For the right hand side
of energy equation we can write
\begin{equation}\label{Q_adv}
    Q_{adv} = Q_{+} - Q_{-} = f Q_{+}
\end{equation}
Here, $ Q_{adv} $ represents the advection transport of energy and is defined
as the difference between the magneto-viscose heating rate and radiative cooling rate.
We employ the advection parameter $ f = 1 - \tfrac{Q_{-}}{Q_{+}} $ to measure
the hight degree to which accretion flow is advection-dominated. When $ f \sim 1 $ the
radiation can be neglected and accretion flow is advection dominated while in the
case of small $ f $, disc is in the radiation dominated case. Now the viscous and
resistive heating rate are expressed as
\begin{equation}\label{dissipation_vis}
    Q_{vis} = \nu \rho r^{2}(\frac{d\Omega}{dr})^{2}
\end{equation}
\begin{equation}\label{dissipation_B}
    Q_{B} = \frac{\eta}{4 \pi}\big|\mathbf{J}\big|^{2}
\end{equation}
where $ \mathbf{J} = \nabla \times \mathbf{B} $ is the current density. We assume both
the kinematic viscosity coefficient $ \nu $ and the
magnetic diffusivity $ \eta $ have the same unites and to be due to turbulence
in the accretion flow. Bisnovatyi-Kogan \& Lovelace 1997 have also shown that the flows have an equipartition magnetic field with the results that dissipation of magnetic energy at a rate comparable to that turbulence must occur by ohmic heating. They argue that this heating occurs as a results of plasma instabilities. Then, it is physically reasonable to express $ \eta $
such as $ \nu $ via the $ \alpha $-prescription of Shakura \& Sunyaev (1973) as
follow (Bisnovatyi-Kogan \& Ruzmaikin 1976)
\begin{equation}\label{nu}
    \nu = \alpha c_{s} H
\end{equation}
\begin{equation}\label{eta}
    \eta = \eta_{0} c_{s} H
\end{equation}
where, $ \alpha $ and $ \eta_{0} $ are the standard viscous parameter and
the magnetic diffusivity parameter respectively. Also these parameters
are assumed to be positive constant
and less than unity $ (\alpha, \eta_{0}\leq 1) $ (Compbell 1999; Kuwabara et al. 2000;
King et al. 2007).

Finally since we consider a global magnetic field, the three components
of induction equation can be written as:
\begin{equation}\label{induction1}
    \dot{B}_{r} = 0
\end{equation}
\begin{equation}\label{induction2}
    \dot{B}_{\varphi} = \frac{d}{dr}\big[(v_{\varphi} B_{r} - v_{r} B_{\varphi})-\frac{\eta}{r}(r B_{\varphi})\big]
\end{equation}
\begin{equation}\label{induction3}
    \dot{B}_{z} = -\frac{1}{r}\frac{d}{dr}\big[r(v_{r} B_{z} + \eta \frac{dB_{z}}{dr})\big]
\end{equation}
where $\dot{B}_{r,\varphi, z}$ are the field scaping/creating rate due to
magnetic instability or dynamo effect. Now we have a set
of MHD equations that describe the structure of magnetized ADAFs.
The solutions to these equations are strongly correlated to
viscosity, resistivity, magnetic field strength $\beta_{r,\varphi,z}$
and the degree of advection $ f $ . We seek a self-similar solution for
the above equations. In the next section we will present self-similar
solutions to these equations.

\section{Self-Similar Solutions}

In order to have a better understanding of the physical
processes taking place in our discs, we seek self-similar
solutions of the above equations. The self-similar method
has a wide range of applications for the full set of MHD
equations although it is unable to describe the global
behavior of accretion flows since no boundary conditions
have been taken into account. However, as long as we
are not interested in the behavior of the flow near the
boundaries, these solutions are still valid. In the self-
similar model the velocities are assumed to be expressed
as follow
\begin{equation}\label{v_r}
    v_{r}(r) = - c_{1} \alpha \sqrt{\frac{G M_{*}}{r_{0}}}(\frac{r}{r_{0}})^{-\frac{1}{2}}
\end{equation}
\begin{equation}\label{v_phi}
    v_{\varphi}(r) =  c_{2} \sqrt{\frac{G M_{*}}{r_{0}}}(\frac{r}{r_{0}})^{-\frac{1}{2}}
\end{equation}
\begin{equation}\label{c_s}
    c^{2}_{s}(r) =  c_{3} (\frac{G M_{*}}{r_{0}})(\frac{r}{r_{0}})^{-1}
\end{equation}
\begin{equation}\label{c_s_rphiz}
    c^{2}_{r,\varphi,z}(r) = \frac{B^{2}_{r,\varphi,z}}{4\pi \rho} =2 \beta_{r,\varphi,z} c_{3} (\frac{G M_{*}}{r_{0}})(\frac{r}{r_{0}})^{-1}
\end{equation}
where constants $ c_{1} $,$ c_{2} $ and $ c_{3} $ are determined later from the
main MHD equations. Also $ r_{0} $ are exploited to write the equation in
non-dimensional form and the constants $ \beta_{r,\varphi,z} $ measure the radio of
the magnetic pressure in three direction to the gas pressure, i.e.,
$ \beta_{r,\varphi,z} = p_{mag,r,\varphi,z}/p_{gas} $.

Assuming the surface density $ \Sigma $ to be in the form of
\begin{equation}\label{self_sigma}
    \Sigma(r) = \Sigma_{0} (\frac{r}{r_{0}})^{s}
\end{equation}
where $ \Sigma_{0} $ and $ s $ are constant. we obtained,e.g.
\begin{equation}\label{self_rhodot}
    \dot{\rho} = \dot{\rho}_{0} (\frac{r}{r_{0}})^{s - \frac{5}{2}}
\end{equation}

\begin{equation}\label{self_Bdot}
    \dot{B}_{r,\varphi,z} = \dot{B}_{r0,\varphi0,z0} (\frac{r}{r_{0}})^{\frac{s - 5}{2}}
\end{equation}
where $ \dot{\rho} $ and $ \dot{B}_{r0,\varphi0,z0} $ are constants.
The half-thickness of the disc still satisfies the relation $ H \propto r $
and we can write
\begin{equation}\label{self_H}
    H(r) = c_{4} r_{0}(\frac{r}{r_{0}})
\end{equation}
By substituting the above self-similar solutions in
the continuity, momentum, angular momentum, hydrostatic
balance and energy equation of the disc, we obtain the following
system of dimensionless equations to be solved for $ c_{1} $,$ c_{2} $,
$ c_{3} $ and $c_{4}$:
\begin{equation}\label{continuityself}
\dot{\rho} = -\big(s + \frac{1}{2}\big)\frac{\alpha c_{1}\Sigma_{0}}{2 r_{0}^{2} c_4}\sqrt{\frac{GM_*}{r_{0}}}
\end{equation}
\begin{equation}\label{motion1}
-\frac{1}{2}\alpha^2 c^2_1 = c^2_2  - 1  - \big[(s - 1)+(s - 1)\beta_z + (s +1)\beta_{\varphi} \big]c_3
\end{equation}
\begin{equation}\label{motion2}
-\frac{1}{2} \alpha c_1 c_2 = -\frac{3}{2} (s + 1) \alpha c_{2} \sqrt{c_{3}} c_{4} + (s + 1) c_3 \sqrt{\beta_r \beta_{\varphi}}
\end{equation}
\begin{multline}\label{moshen3}
    c_{4} = \frac{1}{2}\big[\sqrt{(s - 1)^2 \beta_r \beta_{\varphi} c^2_3 + 4 (1 + \beta_r + \beta_{\varphi}) c_3} \\
    + (s - 1)\sqrt{\beta_r \beta_{\varphi}}c_3\big]
\end{multline}
\begin{multline}\label{energy}
    \big( \frac{1}{\gamma - 1} - (s - 1)\big)\alpha c_1 \sqrt{c_3} = \frac{9}{4} f \alpha  c^2_2 c_4 \\
    + \frac{1}{2} f \eta_{0} c_{3} c_{4} \big[s^{2} \beta_{\varphi} + (s - 2)^{2}\beta_z \big]
\end{multline}
If we solve the self-similar structure of the magnetic field escaping
rate, we obtain
\begin{equation}\label{inducton11}
    \dot{B}_{0r} = 0,
\end{equation}
\begin{multline}\label{inducton12}
    \dot{B}_{0\varphi} = (\frac{s - 3}{2}) \frac{GM_*}{r_{0}^{5/2}}\bigg\{ c_2 \sqrt{\frac{4 \pi  \beta_r c_3 \Sigma_0}{ c_4 }}\\
+ (\alpha c_1 - \frac{s}{2}\eta_{0} \sqrt{c_{3}} c_{4})\sqrt{\frac{4 \pi  \beta_{\varphi} c_3 \Sigma_0}{ c_{4}}} \bigg\}
\end{multline}
\begin{multline}\label{inducton13}
     \dot{B}_{0z}  = (\frac{s - 1}{2}) \frac{GM_*}{r_{0}^{5/2}}\sqrt{\frac{4 \pi  \beta_{z} c_3 \Sigma_0}{ c_4 }}\times\\
\bigg \{ \alpha c_{1} -  (\frac{s - 2}{2})\eta_{0}\sqrt{c_{3}} c_{4}\bigg\}
\end{multline}
It is evident from the above equations that for $ s = -1/2 $
there is no mass loss while it is present for the
case where $ s > -1/2 $. On the other hands, the toroidal component of
escape and creation of magnetic fields balance one another for
$ s = 3 $ and also for $ z $-component of field creating / escaping rate (eq.\ref{inducton13})
they balance for $ s = 1 $.
Although outflow is one of the most important
processes in accretion theory, (see Narayan \& Yi 1995;
Blandford \& Begelman 1999; Stone \& Pringle \& Begelman 1999
and also some recent work like Xie \& Yua 2008;
Ohsuga \& Mineshige 2011), however in our model we choose
a self-similar solution in which $ \dot{\rho} = 0 $, $ (s = -1/2) $ thus
ignoring the effect of wind and outflow on the structure of the disc.

Without magnetic resistivity and magnetic field,
$ \eta=\beta_{r} = \beta_{\varphi} = \beta_{z} = 0 $,
the equations and their similarity solutions are reduced to the standard ADAFs
solution (Narayan \& Yi 1994). Also without resistivity
they are reduced to Zhang \& Dai 2008.

\begin{figure*}
\centering
\includegraphics[width=17cm]{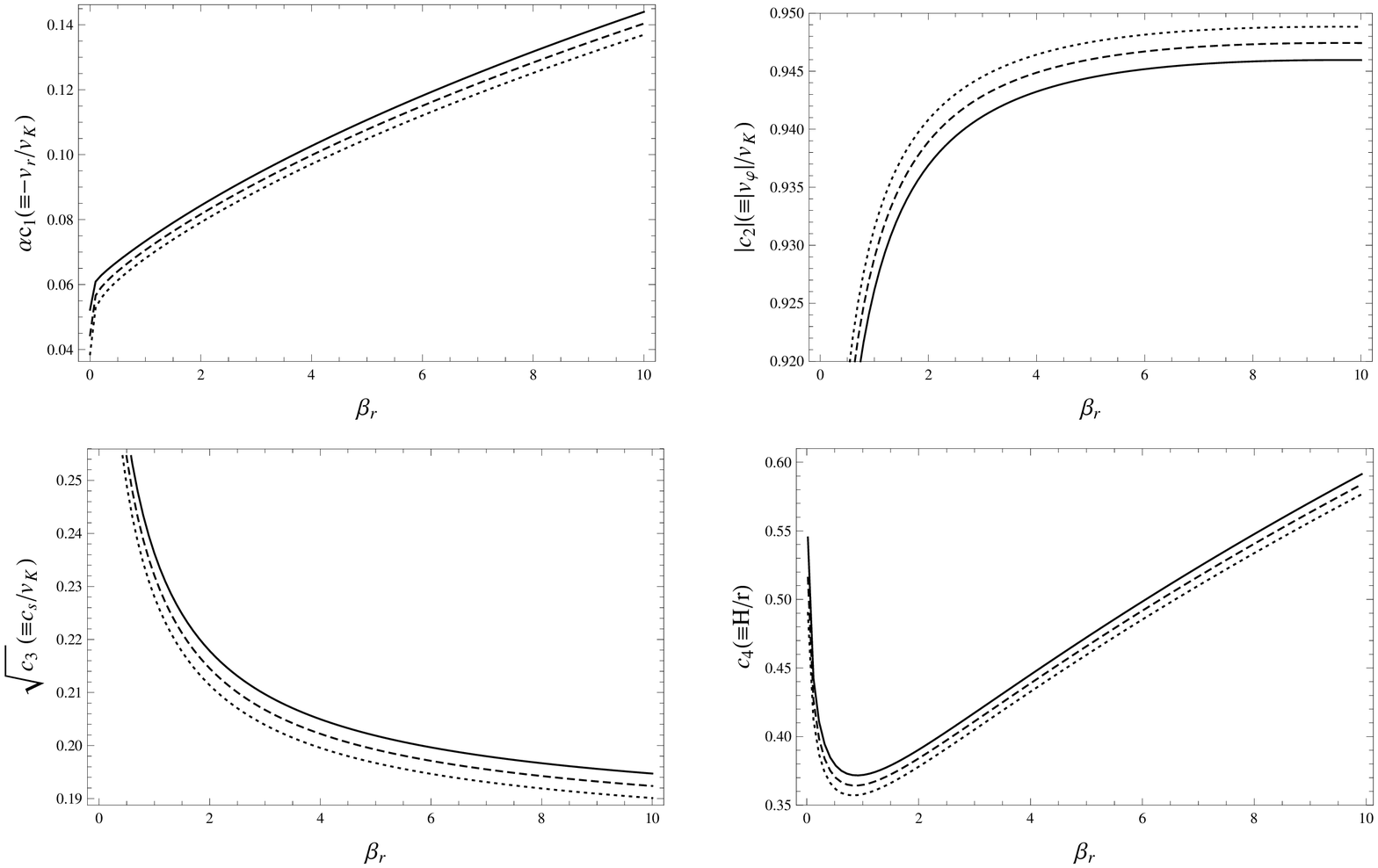}
\caption{Numerical coefficient $ c_{i} $s as a function of magnetic parameter
$ \beta_{r} $ for several values of $ \eta_{0} $. The dotted, dashed and solid
lines correspond to $ \eta_{0} = 0.0, 0.05 $ and $ 0.1 $ respectively. Parameters
are set as $ s = - 0.5 $(no wind), $ \alpha = 0.1 $, $ \beta_{\varphi} = \beta_{z} = 1 $
and $ f = 1 $.}
\label{beta-r}
\end{figure*}

Now, we can do a parametric study considering our input parameters.
The parameters of our model are the standard viscose parameter
$ \alpha $, the magnetic diffusivity parameter $ \eta_{0} $,
the advection parameter $ f $, the radio of the specific heats $ \gamma $ and the
degree of magnetic pressure to the gas pressure in three dimensions of cylindrical coordinate,
$ \beta_{r} $, $ \beta_{\varphi} $ and $ \beta_{z} $. Figure \ref{beta-r}
shows how the coefficients $ \alpha c_{1}(=-v_{r}/v_{K} $, $ c_2(=v_{\varphi}/v_{K}) $,
$ \sqrt{c_{3}} (= c_{s}/v_{K})  $ and $ c_{4}(=H/r) $  depend on the magnetic parameter in radial
direction
$ \beta_{r} $ for several values of magnetic diffusivity parameter $ \eta_{0} $, i. e.
$  \eta_{0} = 0 $  (dotted line) , $  \eta_{0} = 0.05 $  (dashed line) and
$  \eta_{0}= 0.1  $ (solid line), corresponding
to $ \alpha = 0.1 $, $ \gamma = 4/3 $, $ \beta_{\varphi} = \beta_{z} = 1.0 $  and $ f = 1.0 $
(fully advection). radial velocity is shown in the upper-left panel. In ADAFs the radial velocity
is generally less than free fall velocity on a point mass, but it becomes larger if
the magnetic parameter $ \beta_{r} $ is increased. Also, we can see that the radial flows of the
accretion materials become larger by the increase of resistivity parameter $ \eta_{0} $.
The upper right panel of figure \ref{beta-r} shows the radio of the rotational velocity
to the Keplerian one. As it can seen, the rotational velocity slightly shifts up when the magnetic
parameter $ \beta_{r} $ increases. Although the increase of resistivity parameter $ \eta_{0} $ will
decrease the rotational velocity. Moreover, the down left and right panels of figure \ref{beta-r}
display the radio of the sound speed to the Keplerian velocity and the vertical thickness of the disc
respectively. As the magnetic parameter $ \beta_{r} $ become large, the isothermal sound speed of the flow
decreases but vertical thickness of the disc has a different behavior between $ \beta_{r}<1 $ and
$ \beta_{r}>1 $. When $ \beta_{r} $ is below unity, the vertical thickness is reducing. On the other hand,
for $ \beta_{r}>1 $ the thickness is increasing. It means that there is a minimum near
$ \beta_{r} \sim 1 $. Also, by adding the magnetic diffusivity parameter, the sound
speed and vertical thickness will increase.

\begin{figure*}
\centering
\includegraphics[width=17cm]{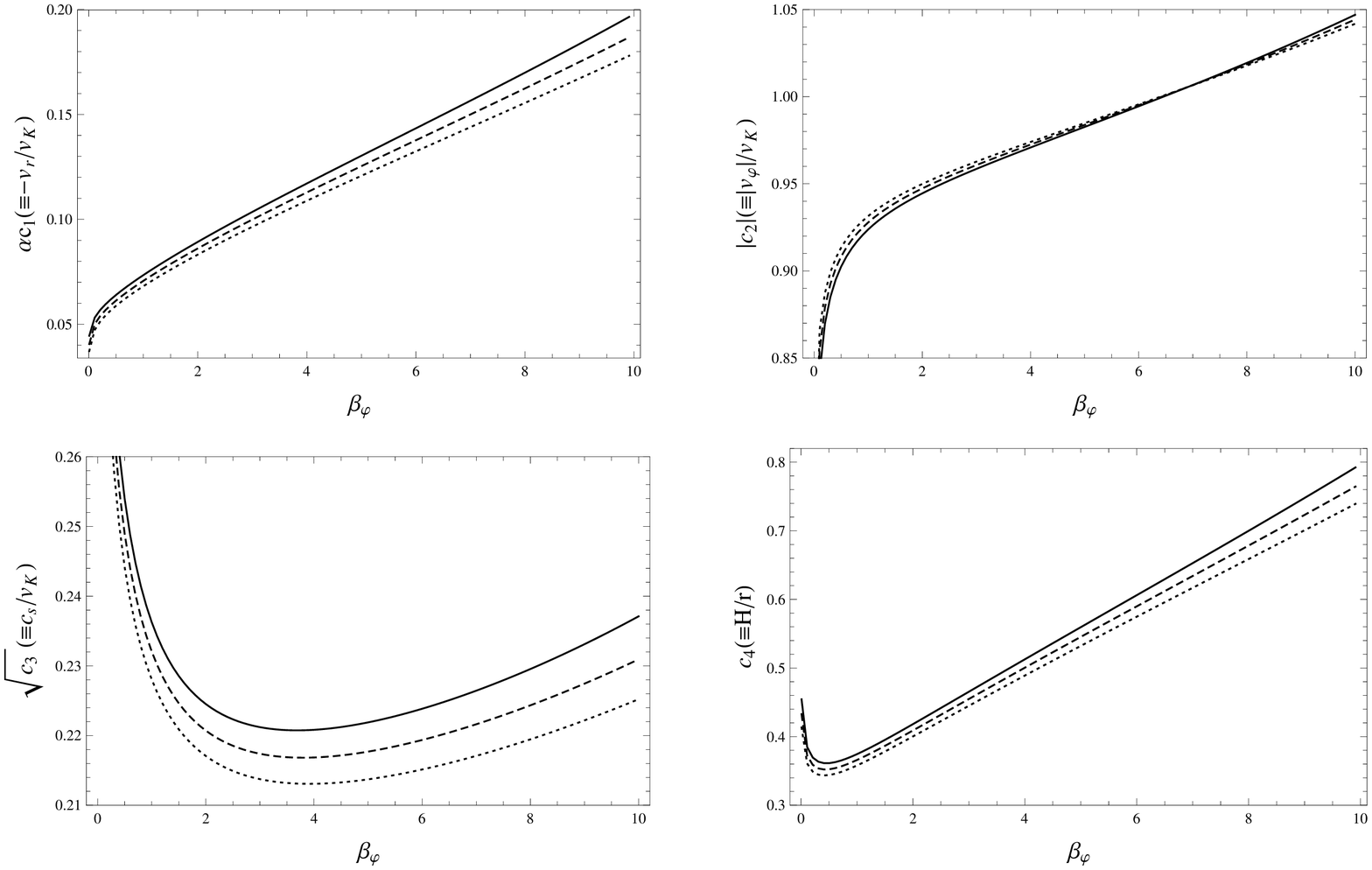}
\caption{Numerical coefficient $ c_{i} $s as a function of magnetic parameter
$ \beta_{\varphi} $ for several values of $ \eta_{0} $. The dotted, dashed and solid
lines correspond to $ \eta_{0} = 0.0, 0.05 $ and $ 0.1 $ respectively. Parameters
are set as $ s = - 0.5 $(no wind), $ \alpha = 0.1 $, $ \beta_{r} = \beta_{z} = 1 $
and $ f = 1 $.}
\label{beta-phi}
\end{figure*}

In figure \ref{beta-phi} behavior of the coefficients $ \alpha c_{1} $, $ c_{2} $, $ c_{3} $
and $ c_{4} $ versus toroidal magnetic field $ B_{\varphi}\ (\beta_{\varphi}) $ are shown
for different values of $ \eta_{0} $. From the upper-left panels of figure \ref{beta-phi},
we can see that for $ \beta_{\varphi} =0-10$, the radial infall velocity is sub-Keplerian, and
that becomes larger by increasing of resistivity parameter $ \eta_{0} $. In addition, when the
toroidal magnetic field becomes stronger, the radial velocity of accretion materials increases.
The upper-right panel in figure \ref{beta-phi} display the rotational velocity of accretion disc.
We see that for the given advection parameter $ f(=1.0) $  and $ \beta_{r} = \beta_{z} = 1.0 $,
the rotational velocity increase as the toroidal magnetic field become stronger. The considerable matter about this panel is that, when  the toroidal magnetic field is very strong, the
rotational velocity will be Super-Keplerian (i.e., $ \beta_{\varphi}>7 $). Also as the
magnetic diffusivity parameter $ \eta_{0} $ increases, the rotational
velocity of the disc decrease. The raise of viscose torque by adding $ \eta_{0} $ and
$ \beta_{\varphi} $ parameters generate a large negative torque in angular momentum equation
and cause the angular velocity of the flow decreases and materials accretes with large speed.
Two down-panels of figure \ref{beta-phi} show the behavior of sound speed and vertical thickness
of the accretion disc respectively. As it can be seen from these panels, by the increase of magnetic diffusivity
($ \eta_{0} $) from $ 0-0.1 $ the isothermal sound speed as well as vertical speed increase. Although we see that
when the toroidal magnetic field increase, at first, the diagrams decrease and then increase. It means that
there is a minimum in two panels.

\begin{figure*}
\centering
\includegraphics[width=17cm]{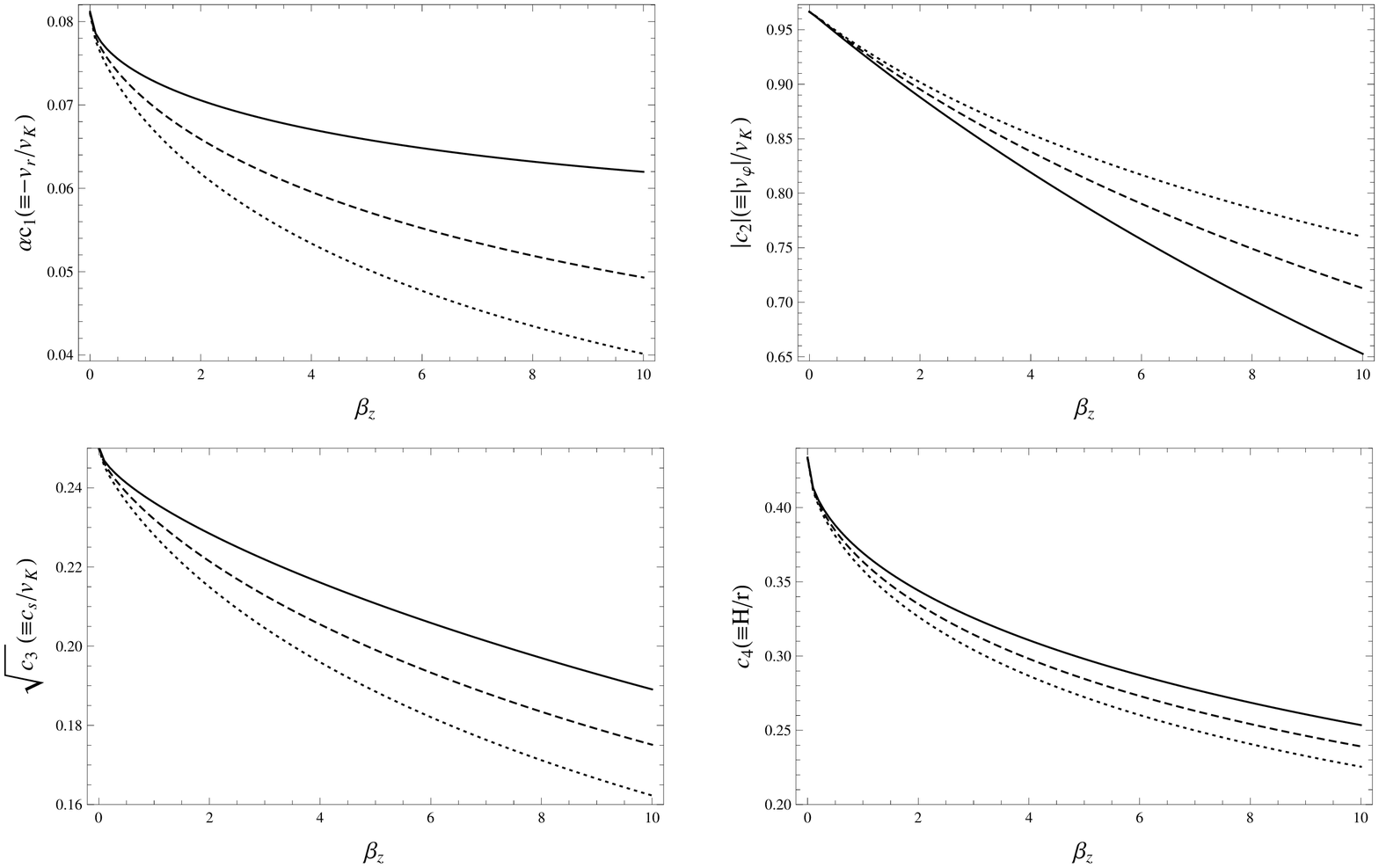}
\caption{Numerical coefficient $ c_{i} $s as a function of magnetic parameter
$ \beta_{z} $ for several values of $ \eta_{0} $. The dotted, dashed and solid
lines correspond to $ \eta_{0} = 0.0, 0.05 $ and $ 0.1 $ respectively. Parameters
are set as $ s = - 0.5 $(no wind), $ \alpha = 0.1 $, $ \beta_{r} = \beta_{\varphi} = 1 $
and $ f = 1 $.}
\label{beta-z}
\end{figure*}

In figure \ref{beta-z} we have plotted the coefficients $ c_{i} $s with $ z $-component of magnetic field
for diffracts values of $ \eta_{0} $. We first find out a strong $ z $-component of magnetic field
$ (\beta_{z}) $ leads to an decrease of the infall velocity $ v_{r} $, rotational velocity $ v_{\varphi} $,
isothermal sound speed $ c_{s} $ and vertical thickness of the disc $ H/r $, which means that a strong
magnetic pressure in the vertical direction prevents the disc matter from being accreted, and decreases
the effect of gas pressure as accretion proceeds. Although it is shown from all four panels  of
figure \ref{beta-z}, the magnetic diffusivity increases $ c_{i} $s except $ c_{2} $ (the rotational velocity).

\section{Summery and Dissussion}

In this paper we studied the influence of the resistivity on the
dynamics of advection dominated accretion flows in the presence
of global magnetic field. Some approximations were made in order
to simplify the main equations. We assumed an axially symmetric,
static disc and the $ \alpha $-prescription is used for the kinematic
coefficient of viscosity $ ( \nu = \alpha c_s H ) $ and the magnetic
diffusivity $ ( \eta = \eta_{0} c_s H ) $. Also we ignored the self-gravity
and the relativistic effects.

We have extended Akizuki \& Fukue 2006 and Zhang \& Dai 2008 self-similar solutions to
present dynamical structure of advection dominated accretion flow by considering
a three component magnetic field and magnetic diffusivity.
We have accounted for this possibility by allowing input parameters
$ \alpha $, $ \eta_{0} $, $ \beta_{r} $, $ \beta_{\varphi} $ , $ \beta_{z} $, $ f $
to vary in our solutions.

First, we have shown that the physical quantities of the disk are sensitive to the magnetic
diffusivity parament $ \eta_{0} $. We found out the magnetic resistivity increase
the radial infall velocity $ v_{r} $ , the isothermal sound speed $ c_{s} $ and the vertical thickness of
the disk $ H/r $.
On the other hand, our numerical results shown that  when the magnetic
diffusivity parameter raise up, the rotational velocity of the disk $ | v_{\varphi} | $ decrease.

Also we have shown that the strong magnetic filed in the radial direction increase
$ v_r $, $ | v_{\varphi} | $ and $ H/r $ although $ c_{s} $ will be decrease which
are satisfied with the results were presented by Zhang \& Dai 2008. Also when the
toroidal magnetic field become stronger, our physical quantities  $ v_r $, $ | v_{\varphi} | $, $ c_{s} $ and
$ H/r $ increase (see, e.g., Akizuki \& Fukue 2006; Abbassi et al 2008, 2010; Bu \& Yuan \& Xie 2009).
Moreover according to Zhang \& Dai 2008, a large vertical magnetic field prevents the disk from
being accreted and decrease effect of the gas pressure.

Although our preliminary self-similar solutions are too simplified, they clearly improve
our understanding of the physics of ADAFs around a black hole. In order to have a realistic picture of an
accretion flow a global solution is needed rather than the self-similar one. In our future studies
we intend to investigate the effect of wind and thermal conduction on the observational appearance and
properties of a hot magnetized flow.

\acknowledgments

The authors would like to thank the anonymous referee for his/her helpful suggestions and
careful reading, which improve the manuscript. This work has been supported financially by Center for Excellence in Astronomy \& Astrophysics
(CEAA-RIAAM) under research project No. 1/2643.

\end{document}